\documentclass[twocolumn]{aastex631}

\usepackage{placeins}
\usepackage{xspace}
\usepackage{natbib}
\bibpunct{(}{)}{;}{a}{}{,}
\usepackage{graphicx}
\usepackage{xcolor}
\usepackage{multirow}
\usepackage{amsmath}
\usepackage{lipsum}
\usepackage{calrsfs}

\newcommand{\NH}{\ensuremath{N_{\mathrm{H}}}\xspace}

\newcommand{\xabs}{{\tt xabs}\xspace}
\newcommand{\amol}{{\tt amol}\xspace}
\newcommand{\hot}{{\tt hot}\xspace}

\newcommand{\bb}{{\tt bb}\xspace}

\newcommand{\gau}{{\tt gau}\xspace}

\newcommand{\spex}{{\tt SPEX}\xspace}
\newcommand{\rslmkrmf}{{\tt rslmkrmf}\xspace}
\newcommand{\xaarfgen}{{\tt xaarfgen}\xspace}

\newcommand{\feix}{\ion{Fe}{9}\xspace}
\newcommand{\caix}{\ion{Ca}{9}\xspace}
\newcommand{\fei}{\ion{Fe}{1}\xspace}
\newcommand{\feii}{\ion{Fe}{2}\xspace}
\newcommand{\feiii}{\ion{Fe}{3}\xspace}
\newcommand{\feiv}{\ion{Fe}{4}\xspace}

\begin{document}

% -------------- Affiliation and title ------------ %
%`\title{Iron in interstellar dust: XRISM Fe K-edge absorption spectrum of GX 340+0}
\title{A first view of interstellar Fe-bearing dust grains with XRISM}

\author[0000-0002-1049-3182]{I. Psaradaki}
\affiliation{European Space Agency (ESA), European Research and Technology Centre (ESTEC), Keplerlaan 1, 2201 AZ Noordwijk, The Netherlands}
\affiliation{MIT Kavli Institute for Astrophysics and Space Research, 70 Vassar Street, Cambridge, MA 02139}

\author{L. Corrales}
\affiliation{Department of Astronomy, University of Michigan, 1085 S. University, MI 48109, USA}

\author{E. Costantini}
\affiliation{SRON Space Research Organisation Netherlands, Niels Bohrweg 4, 2333 CA Leiden, The Netherlands}
\affiliation{Anton Pannekoek Institute for Astronomy, University of Amsterdam, Science Park 904, NL-1098 XH Amsterdam, The Netherlands}

\author{M. Mehdipour}
\affiliation{Department of Astronomy, University of Michigan, 1085 S. University, MI 48109, USA}

\author{D. Rogantini}
\affiliation{Department of Astronomy and Astrophysics, University of Chicago, 5640 S Ellis Ave, Chicago, IL 60637, USA}

\author{H. Takahasi}
\affiliation{Department of Physics, Hiroshima University, Hiroshima 739-8526, Japan}

\author{R. Ballhausen}
\affiliation{Department of Astronomy, University of Maryland, College Park, MD 20742, USA}
\affiliation{NASA / Goddard Space Flight Center, Greenbelt, MD 20771, USA}

\author{P. Chakraborty}
\affiliation{Department of Physics, University of Arkansas, 825 W Dickson st., Fayetteville, AR, 72701, USA}
\affiliation{Center for Astrophysics — Harvard \& Smithsonian, Cambridge, MA, USA}

\author{T. Nakagawa}
\affiliation{Advanced Research Laboratories, Tokyo City University, 1-28-1 Tamazutusumi, Setagaya-ku, Tokyo 158-8557, Japan}
\affiliation{Institute of Space and Astronautical Science, Japan Aerospace Exploration Agency, 3-1-1 Yoshinodai, Chuo-ku, Sagamihara, Kanagawa 252-5210, Japan}

\author{R. K. Smith}
\affiliation{Center for Astrophysics — Harvard \& Smithsonian, Cambridge, MA, USA}

\author{S. Zeegers}
\affiliation{SRON Space Research Organisation Netherlands, Niels Bohrweg 4, 2333 CA Leiden, The Netherlands}
\affiliation{Anton Pannekoek Institute for Astronomy, University of Amsterdam, Science Park 904, NL-1098 XH Amsterdam, The Netherlands}

\begin{abstract}

Understanding the reservoirs of iron (Fe) in our Galaxy is crucial for probing the mechanisms that drive dust grain growth in the interstellar medium (ISM). High-resolution X-ray absorption spectroscopy offers a powerful tool to investigate the chemistry of dust grains, particularly those containing Fe, through spectroscopic features near photoabsorption edges known as X-ray Absorption Fine Structures (XAFS). With the Resolve spectrometer aboard XRISM, we can now examine the Fe K-edge ($\rm \sim 7.1 \ keV$) owing to its improved spectral resolution in this energy range. This enabled us to probe the dust grain chemistry in high Galactic column densities. In this work, we present a pilot study on dust grain chemistry through the Fe K-edge of the XRISM Performance Verification target GX 340+0. This source is a persistently bright low-mass X-ray binary with a hydrogen column density of $\rm \sim 6.6\times 10^{22} \ cm^{-2}$ along the line of sight, providing sufficient optical depth at the Fe K-edge. By fitting combined dust and gas models, we find that Fe is depleted by $\sim$95\% into dust along the line of sight towards the source. Moreover, we find evidence that $\sim$38\% of the solid-phase Fe is in metallic iron, $\sim$47\% in iron sulfides, and the remainder in silicates. The large contribution from metallic iron and iron sulfides differs from what is typically inferred from
X-ray studies of diffuse ISM and may suggest that that these compounds
become more significant reservoirs of Fe in higher column density Galactic sightlines.

\end{abstract}

\section{Introduction} \label{sec:intro}

%%%%%%%%%-----------------------------------------------%%%%%%%%%%%%
%%%%%%%%%--------------- Introduction ------------------%%%%%%%%%%%%
%%%%%%%%%-----------------------------------------------%%%%%%%%%%%%

Interstellar dust grains are small particles (typically ranging in size from a few nanometers to a few micrometers) and constitute an important component of the interstellar medium in galaxies. Dust grains cool down interstellar clouds, enabling star formation, provide the surface for chemical reactions, and are the building blocks from which planetesimals are formed in planet-forming disks around young stars \citep{Henning2010}. Dust, mainly characterised by its chemical composition, size and crystallinity (i.e. lattice structure), contains much of the elements, such as Fe, O, Si, Mg, S and C, in solid form. Understanding the composition, origin, and evolution of interstellar dust is a key question in astrophysics \citep{bookWhittet, Tielensbook,Drainebook}.

Identifying the composition of solid-phase iron is crucial for unraveling the processes of dust grain growth in the interstellar medium (ISM), which has been a puzzle in dust studies throughout the years. Iron constitutes a major component in most dust grain models. Unlike other refractory elements like Mg and Si, which are primarily synthesized in stars with dust-forming outflows, Fe is predominantly synthesized in Type Ia supernovae. Approximately 70\% of newly synthesized Fe is introduced into the ISM in the gas phase (\citealt{Dwek2016}). Intriguingly, astronomical observations indicate that over 90\% of Fe is depleted from the gas phase in the ISM (e.g., \citealt{Savage1979}, \citealt{Jenkins1986}, \citealt{Snow2002}, \citealt{Jenkins2009}), implying that Fe must condense within the ISM. However, the mechanism and the interstellar environment for this condensation remain unknown.

Silicates are minerals composed primarily of silicon, oxygen, and metals such as magnesium and iron, and they form a major component of interstellar dust \citep{Henning2010}. They play a critical role in the life cycle of matter in the ISM, influencing its chemistry, thermal structure, and providing surfaces for chemical reactions. Pyroxenes, (Mg, Fe)SiO$_3$, and olivines, (Mg, Fe)$_2$SiO$_4$, are the two main silicate groups expected to describe the chemistry of interstellar dust, but the exact ratio of olivines to pyroxenes, as well as the Mg:Fe ratio, remain uncertain. Understanding this composition is important, as Mg-rich silicates may explain the high magnesium abundance observed in cometary and circumstellar grains \citep{Min2007, Fogerty2016}. 
Evidence suggests a departure from the co-condensation of Fe with Mg and Si, pointing towards a non-silicate host for Fe in the solid phase \citep{Dwek2016}. Additional iron reservoirs, beyond silicates, have been discussed in the literature. Ferromagnetic inclusions ($\rm Fe_{3}O_{4}$) could be possible iron carriers. In particular, dust grain inclusions of ferromagnetic material, such as magnetite, have been discussed in the view of grain alignment due to their ferromagnetic properties (\citealt{Lazarian}). FeS compounds are also potential candidates of interstellar dust \cite[e.g.][]{Wooden2008,Westphal2014}. These compounds are found in meteorites, interplanetary dust particles, and some cometary dust particles \cite[e.g.][]{Bradley2010, HannerZolensky2010, MannJessberger2010}. 

X-ray absorption studies of the diffuse ISM \cite[e.g.][]{Psaradaki2022,Corrales2024,Psaradaki2026} found only a negligible amount of dust column density to exist in either sulfides (FeS and Fe$_2$S) or iron oxides (magnetite, Fe$_3$O$_4$). \cite{Westphal} proposed that Fe is distributed roughly equally between sulfide and metallic phases, consistent with a high-temperature condensation sequence in which Fe and S are bound to these components. 

\citet{Schalen1965} and \citet{Kemper2002} suggested that a portion of the depleted cosmic iron could be in the form of pure metallic iron. Metallic iron, typically found in solid phase, is of prime importance in the exploration of interstellar dust chemistry, attracting considerable attention as a potential carrier for iron-bearing dust. 
Spectroscopic studies, in different wavebands, from the infrared to X-rays have found that the spectral profile shapes align more closely with Mg-rich rather than Fe-rich silicates (\citealt{Min2007}, \citealt{vanBreemen_2011}, \citealt{Poteet}, \citealt{Psaradaki2022}). This leaves uncertainty about the specific form in which the majority of solid-phase Fe is present allowing for the possibility of metallic iron. 

\cite{Poteet} investigated the chemical composition of interstellar dust towards $\zeta$ Ophiuchi, finding that 40\% of the iron exists in the form of silicate grains. The rest could be present in various forms such as metallic nanoparticles, Fe oxides, or metallic inclusions in glass with embedded metal and sulfides of interstellar origin (GEMS, e.g., \citealt{Bradley1994}, \citealt{Keller}). Moreover, hydrodynamic simulations of \citet{Zhukovska2018} suggest that a large fraction of iron ($\rm 70\%$) is locked as inclusions in silicate grains, where it is protected from sputtering by SN shocks. The remaining depleted iron could reside in a population of metallic iron nanoparticles with sizes in the range of 1–10 nm.

High-resolution X-ray spectroscopy is a powerful tool for studying the interstellar medium (e.g. \citealt{Wilms, Ueda2005, Juett2004, Paerels, Smith2008, García2011, Pinto2013, Costantini2012, Schulz, Gatuzz2016, Corrales2016, Zeegers2019, Rogantini2019, Ballhausen, Psaradaki2020, Yang2022, Moutard2026}). In particular, X-rays enable the study of the solid phase composition of highly depleted elements such as iron in the line of sight towards a bright background source. X-ray absorption fine structures (XAFS) are spectroscopic features observed near the photoelectric absorption edges of solid material (dust), and their shape is the ultimate footprint of the dust chemical composition \cite[e.g.][]{Newville, Lee2005, Zeegers2019, Rogantini2018, Rogantini2019, Westphal, Costantini2022, Psaradaki2022}. 

Iron exhibits two types of absorption edges in the X-ray spectrum: the Fe K-edge at 7.112~keV and three Fe L-edges at 0.846, 0.721, and 0.708~keV (L\textsubscript{I}, L\textsubscript{II}, L\textsubscript{III}). The Fe K-edge serves as a probe of high Galactic column densities, as these are required to achieve sufficient optical depth in this energy range compared to the Fe L-edges. However, this energy range has been less extensively explored due to the limited spectral resolution available in previous X-ray telescopes \citep{Rogantini2018}. With a spectral resolution of 5~eV, the Resolve instrument on board XRISM \citep{XRISM_white_paper} surpasses earlier X-ray instruments.
The advent of XRISM now enables us to investigate the chemistry of interstellar dust in high Galactic column density environments, particularly focusing on Fe-bearing dust grains.

Laboratory studies have highlighted the potential of constraining the chemical composition of dust grains through the Fe K-edge. Using laboratory data of ISM dust analogs and synchrotron facilities, \cite{Lee2005} demonstrated that, with sufficient signal-to-noise, high-resolution X-ray spectroscopy can distinguish between gas- and dust-phase absorption and even identify specific dust chemistry. More recently, \cite{Rogantini2018} employed synchrotron Fe K-edge data, from a large laboratory campaign \citep{CostantiniDeVries2013}. They showed that the post-edge fine structures trace dust composition, while pre-edge scattering constrains grain size, underscoring the Fe K-edge as a powerful probe of Fe-bearing dust in dense Galactic environments.

Here, we present a pilot analysis of the Fe K edge with XRISM data, focusing on ISM absorption along the line of sight to the XRISM Performance Verification (PV) target GX 340+0.
 GX 340+0 is a neutron-star, low-mass, persistently bright X-ray binary located near the Galactic Plane. It is located at a distance of $11\pm3$ kpc \citep{Penninx1993,Fender2000} with an inclination angle of $20^{o}\leq i \le 41^{o}$ \citep{Fender2000,DAi2008,Cackett2010,Miller2016}. It was classified as a ``Z" source based on its X-ray color-color diagram \citep{Hasinger1989}. The system has been extensively studied in X-rays, primarily to investigate its emission and absorption features \citep[e.g.,][]{Ueda2005, DAi2008, Cackett2010, Miller2016, Ludlam2025, Corrales2025, Chakraborty2025}.

In this work, we focus on the intervening ISM along the line of sight, in particular on Fe-bearing dust grains. Its high Galactic column density \cite[$\rm 6.6 \times 10^{22}\ cm^{-2}$;][]{Zeegers2019, Seifina} and high luminosity \cite[$\rm >0.5 L_{Edd}$,][]{Seifina}, where $L_{Edd}$ is the Eddington luminosity, make GX 340+0 an exceptional target for studying the Fe K-edge with the XRISM/Resolve long exposure time of $\sim$300 ks.
The paper is organized as follows: Section \ref{reduction} describes the data and reduction process, Section \ref{fitting} outlines our spectral fitting approach for the ISM, and in Section \ref{Discussion} we discuss the results.

%%%%%%%%%-----------------------------------------------%%%%%%%%%%%%
%%%%%%%%%--------------- Data reduction ----------------%%%%%%%%%%%%
%%%%%%%%%-----------------------------------------------%%%%%%%%%%%%

\section{Data and reduction} \label{reduction}

Two observations of GX~340+0 were carried out during the XRISM Performance Verification phase: a 152 ks exposure obtained between 17 and 21 August 2024 (ObsID 300002010), hereafter referred to as observation 1, and a 153 ks exposure obtained between 31 August and 4 September 2024 (ObsID 300002020), hereafter as observation 2. These observations have been taken with the Gate Valve closed \footnote{\url{https://heasarc.gsfc.nasa.gov/docs/xrism/proposals/POG/Introduction.html}}.  

In this analysis, we exclusively use observations obtained with the Resolve instrument. The data were reprocessed using Heasoft-6.34, incorporating calibration files from CalDB v8.
Following the Quick-Start Guide \footnote{\url{https://heasarc.gsfc.nasa.gov/docs/xrism/analysis/quickstart/index.html}}, we applied rise-time screening with "STATUS[4]==b0". We then extracted and analyzed only the highest-resolution (Hp) events. The RMFs and ARFs for a point source were generated using the tools \rslmkrmf and \xaarfgen. 

%%%%%%%%%-----------------------------------------------%%%%%%%%%%%%
%%%%%%%%%--------------- Spectral fitting --------------%%%%%%%%%%%%
%%%%%%%%%-----------------------------------------------%%%%%%%%%%%%

\section{Spectral fitting of the Resolve spectrum} \label{fitting}

\begin{figure}
    \centering
    \includegraphics[width=0.5\textwidth]{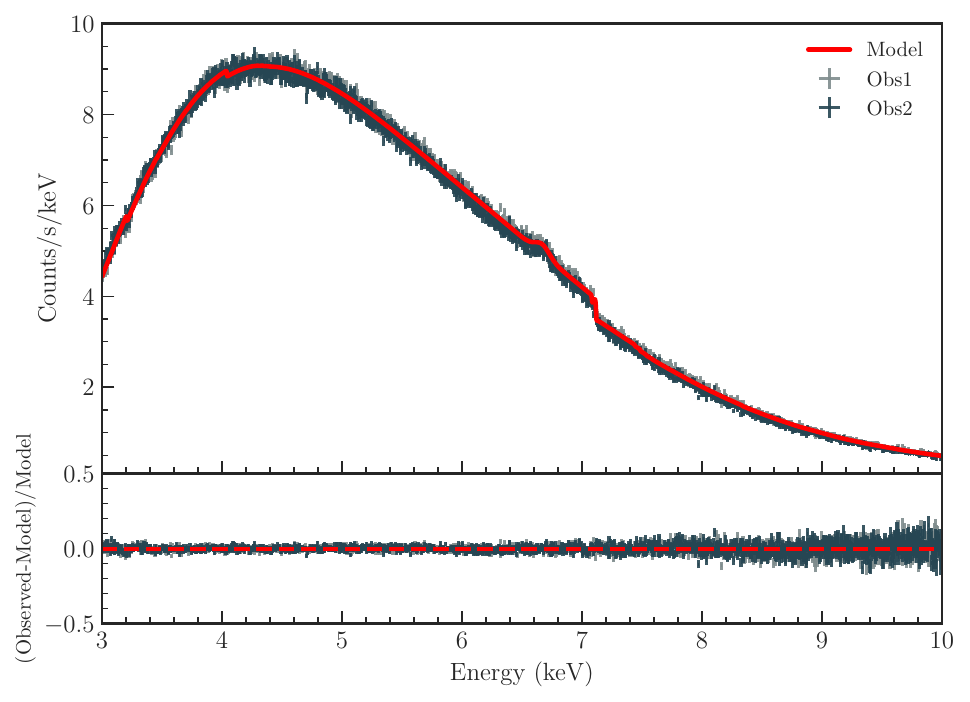}
    \caption{XRISM/Resolve spectrum of GX 340+0 and the best fit model in the 3-10 keV band.}
    \label{fig:continium}
\end{figure}

For our modeling, we use the SPEctral X-ray and UV modeling and analysis software, \spex, version 3.08.01\footnote{\url{http://doi.org/10.5281/zenodo.12771915}} (\citealt{Kaastra1996, kaastra_2024}). To evaluate the goodness of our fits, we employ $C$-statistics, referred to as $C_{\rm stat}$ throughout this manuscript. \spex implements $C_{\rm stat}$ as outlined in \citet{Cash1979}, with an additional component that provides an estimate to the goodness of the fit (see \citealt{Kaastra2017} for details). The errors are provided at 1$\sigma$ (68\%) confidence level. Furthermore, we adopt proto-Solar abundance values from \citet{Lodders2009}.

\newpage
\subsection{Continuum model}

To obtain the continuum parameters, we fit the XRISM/Resolve spectrum in the 3--10 keV energy range. We model the continuum using a thermal blackbody (\bb) component. The corresponding temperature is reported in Table~\ref{tab:parameters}. The temperature is fitted independently for each dataset; however, we obtain consistent values within the uncertainties. Similar models have been employed in previous studies of GX~340+0 \cite[e.g.,][]{DAi2008, Rogantini2026}, and by \citet{Ludlam2025}, who analyzed the source using XRISM/Resolve, NICER, and NuSTAR data. In our case, the fit to the continuum is restricted to a limited portion of the X-ray spectrum (3-10 keV), making the continuum description somewhat phenomenological, although the derived parameters remain physically reasonable.

\citet{Ludlam2025,Chakraborty2025} analyzed the XRISM/Resolve spectrum of this source, focusing on the structure of the relativistic Fe emission line as well as the ionization and velocity structure of the surrounding material. These features primarily affect the spectrum around 6.7 keV, sufficiently separated from the Fe K absorption edge at 7.1 keV, where the dust signatures are expected to appear. Nevertheless, an accurate description of the continuum is important for reliable modeling of the dust absorption edge. We therefore account for these spectral features using both a narrow and a broad gaussian emission component (\gau in \spex). Separate gaussian components are used for each observation, with the parameters fitted independently. A detailed physical interpretation of this spectral region is beyond the scope of the present study, which focuses on the dust absorption edge itself.

To account for possible ionized absorption along the line of sight or in the circumbinary material, we use the \xabs model in \spex. This model calculates the transmission through a slab of material, where all ionic column densities are linked through a photoionization model. We use separate \xabs components for the two observations, with the corresponding hydrogen column density and ionization parameter left free to vary. This model improves the fit to the absorption, but also predicts absorption features from highly ionized species, such as \feix and \caix. Evidence for weak photoionized absorption was also reported by \citet{Chakraborty2025}. Including the components described above, we obtain an adequate description of the continuum, yielding a fit statistics of $C_{\rm stat}/\mathit{d.o.f.}=1538/1472$, where $\it{d.o.f.}$ is the number of degrees of freedom in the fit.
The best-fit parameters are listed in Table~\ref{tab:parameters}.

\begin{table}
\caption{Best-fit parameters.}
\label{tab:parameters}
\scalebox{0.85}{
\begin{tabular}{c c c c}
\hline
\hline
Component & Parameter & Obs. 1 & Obs. 2 \\
\hline\hline

\bb & $kT_{\rm bb}$ & \multicolumn{2}{c}{$1.21\pm0.02$}   \\
\hline

\multirow{3}{*}{\gau}
  & e  & $6.668\pm0.008$ & $6.687\pm0.007$ \\
  & FWHM  & $0.10\pm0.02$ & $0.09\pm0.02$ \\
\hline

\multirow{3}{*}{\gau}
  & e  & $6.70\pm0.02$ & $6.695\pm0.018$ \\
  & FWHM  & $0.75\pm0.07$ & $0.71\pm0.07$ \\
\hline

\multirow{2}{*}{\hot}
  & $N_{\rm H}$ & \multicolumn{2}{c}{$6.9\pm4.5$} \\
  & $kT$        & \multicolumn{2}{c}{$10^{-6}$, fixed} \\
\hline

\multirow{2}{*}{\xabs}
  & $N_{\rm H}$ & $1.6\pm1.1$ & $0.6\pm0.1$ \\
  & $\log\xi$   & $0.5\pm0.3$ & $0.8\pm0.2$ \\
\hline\hline
\end{tabular}}
\begin{minipage}{1\linewidth}
\vspace{0.1cm}
\footnotesize
$kT_{\rm bb}$ denotes the blackbody temperature in keV. The symbol \NH\ represents the hydrogen column density in units of $10^{22} \mathrm{cm^{-2}}$, while $kT$ is the plasma temperature in keV. The quantity $\log \xi$ is the base-10 logarithm of the ionisation parameter $\xi$, expressed in units of $10^{-9} \mathrm{Wm}$. For Gaussian components, we report the line centroid energy (e) and the full width at half maximum (FWHM) in keV. 
\end{minipage}
\end{table}

\subsection{Spectral fitting of the ISM}
To characterize the neutral Galactic absorption, we adopt the \hot model in \spex (\citealt{dePlaa}). For a given temperature and set of elemental abundances, this model calculates the ionization balance and derives the ionic column densities by scaling them to the specified total hydrogen column density. By setting the temperature to a low value ($\sim$0.001 eV, corresponding to $\sim$10 K), the \hot model effectively describes absorption by neutral gas.
The hydrogen column density ($N_{\rm H}$, in $\rm cm^{-2}$) is left free to vary. The gas-phase abundances of Si, Mg, O, and S are fixed to the values reported by \cite{Zeegers2019}, \cite{Rogantini2019}, \cite{Psaradaki2020}, and \cite{Rogantini2026}, respectively, while the Fe abundance is left free to vary. The parameters of the spectral fit are summarized in Table~\ref{tab:parameters}, and the best-fit model is shown in Figure~\ref{fig:continium}.

\begin{figure}
    \centering
    \includegraphics[width=0.48\textwidth]{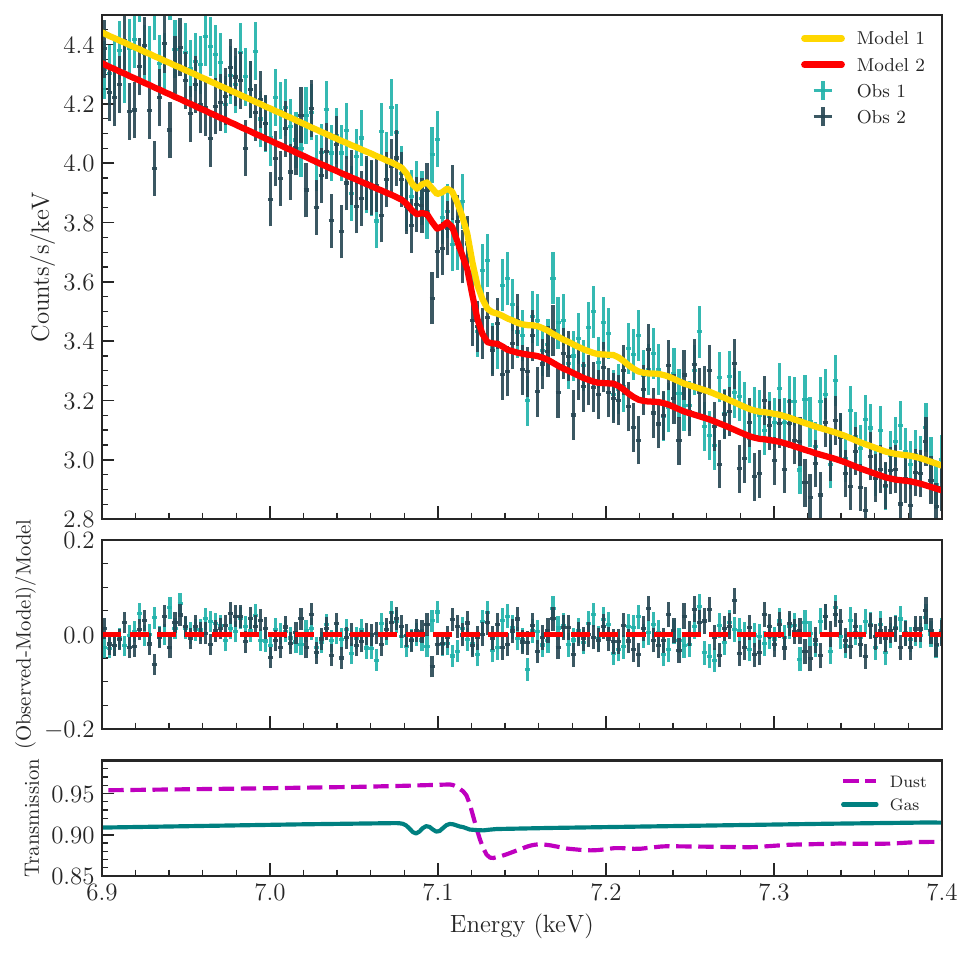}
    \caption{Zoom into the Fe K-edge and the best fit model. Each dataset corresponds to a different observation, and the associated best-fit models are shown as Model 1 and Model 2. The middle panel shows the residuals of the fit, and the bottom panel the transmission of the gas and dust components.}
    \label{fig:FeK}
\end{figure}

To model the dust absorption, we focus on the Fe K-edge region, specifically the 6--7.5 keV energy range. While fitting the full XRISM spectrum is essential to obtain a reliable continuum description, the dust analysis is restricted to the region containing the Fe K-edge feature. We adopt this approach because the GX~340+0 spectrum contains several complex spectral features, including the S K-edge region, which has already been investigated independently by \citet{Corrales2025}. In this pilot study, we specifically focus on the Fe K-edge to characterize the absorption signatures of Fe-bearing dust grains and determine the dust mixture that best describes the data. Therefore, the dust models described below are tested only within the Fe K-edge region rather than across the full spectrum, avoiding additional complexity.

The presence of dust modifies the shape of the photoabsorption edge, providing a diagnostic signature in the X-ray spectrum. To model the dust we employ the \amol\ model in \spex, which calculates the transmission of a dust component while treating the dust column density as a free parameter. For this purpose, we adopt the dust extinction cross sections presented by \citet{Rogantini2018}, which are based on laboratory measurements; a detailed description of the models is provided therein. The scattering halo's contribution to high-resolution spectral features depends on the telescope's imaging resolution \citep{Corrales2016}, and at 7 keV, a fraction of the source's scattering halo is expected to fall within the Resolve field of view. Since the Fe K scattering cross-section is relatively flat compared to the absorption features \citep{Rogantini2018}, any such residual is unlikely to significantly alter the shape of the extinction features.

The dust extinction cross sections were calculated from optical constants derived from laboratory measurements, using Anomalous Diffraction Theory \citep[ADT;][]{van_de_Hulst}, assuming a Mathis-Rumpl-Nordsieck (MRN) dust size distribution \citep{mathis}. This distribution follows a power-law of the form $dn/da \propto a^{-3.5}$, where $a$ represents the grain size, with a lower cutoff at $\rm 0.005 \mu m$ and an upper limit of $\rm 0.25 \mu m$. In Table \ref{tab:samples} we summarize the different types of dust minerals used in this study. The models include different types of silicates (olivine and pyroxenes), iron sulfides (FeS, $\rm Fe_{7}S_{8}$) and metallic iron.

The \amol model can simultaneously fit up to four dust compounds in a single run, so we explore all possible combinations of four species from the available samples, following the assumption that interstellar dust can be represented by at most four components \cite[see also,][]{Costantini2012}. Therefore, we performed a series of independent spectral fits, each corresponding to a unique combination of four dust species. The total number of mixtures is determined by the formula $C_{e,c}=e!/c!(e-c)!$, where $e$ is the number of available edge profiles, 125 in this case, and $c$ the chosen combination class.

\begin{figure*}
    \centering
    \includegraphics[width=\textwidth]{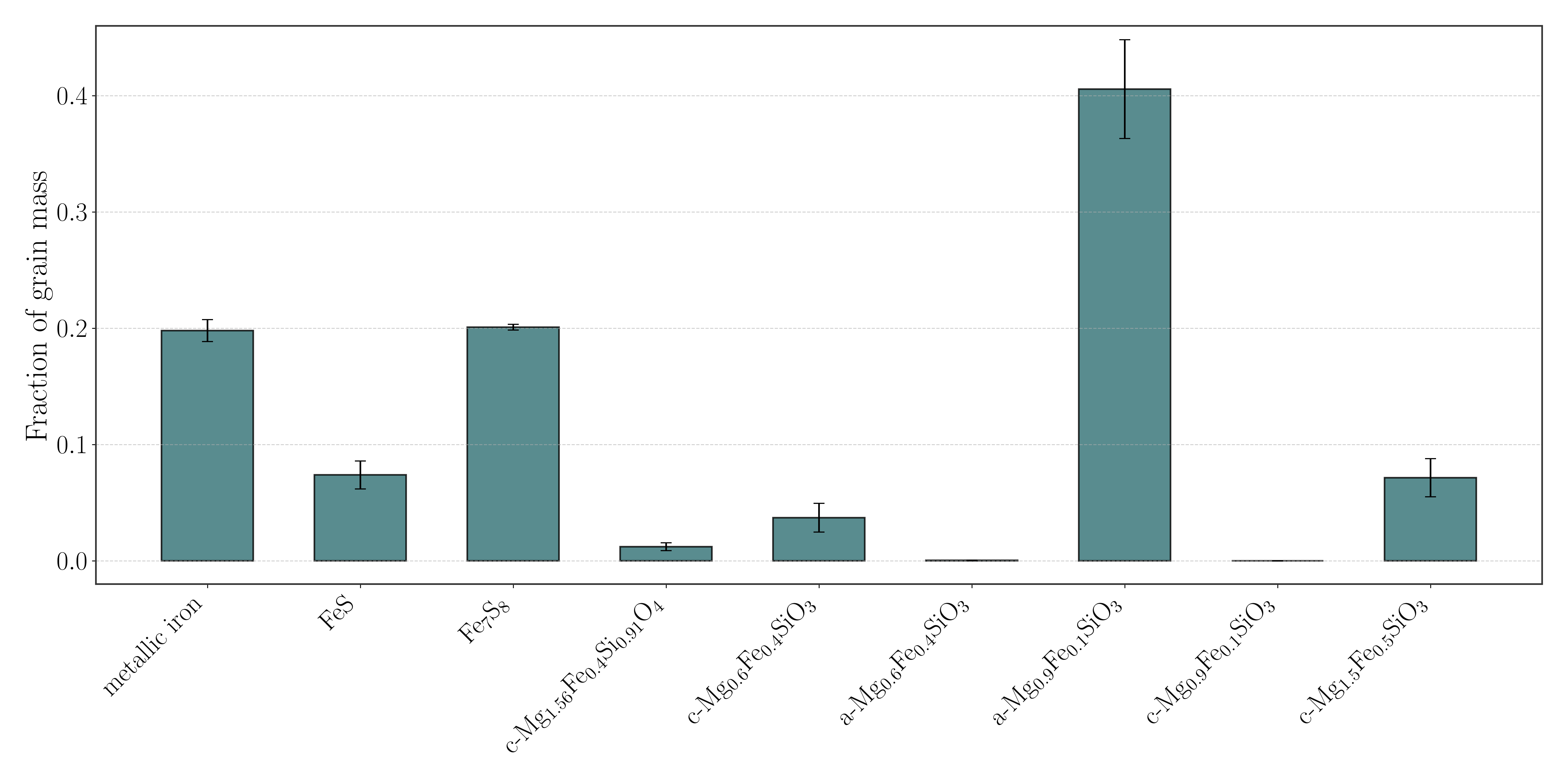}
    \caption{Relative fraction of column density for each dust compound. The fractions have been calculated considering models with $\Delta \mathrm{AIC} < 4$. The symbol $a$ denotes amorphous compounds, while $c$ refers to crystalline ones. The error bars indicate the standard deviation of each fraction within the $\Delta \mathrm{AIC} < 4$ selected models.}
    \label{fig:barplot}
\end{figure*}

Each combination is fitted separately to the data, yielding a distinct set of best-fit parameters and fit statistics. To statistically evaluate which mixture provides the best description of the spectra, we use the \textit{Akaike Information Criterion} (AIC; \citealt{Akaike}), which has been successfully applied in dust studies before (e.g., \citealt{Rogantini2019}). AIC provides a robust and efficient way to compare models without requiring the assumption that any candidate model is the true one. It is defined as:

\begin{equation}
\mathrm{AIC} = 2k - 2 \ln(\Lambda_{\max}),
\end{equation}

\noindent where $k$ is the number of free parameters in the model and $\Lambda_{\max}$ is the maximum likelihood value. Since the C-statistic is related to the likelihood via $\mathrm{Cstat} = -2 \ln(\Lambda)$ \citep{Cash1979}, the AIC effectively allows a quick comparison of the candidate's models by comparing the C-statistic value of every fit with the best one. In practice, we ranked all models according to their AIC values and computed the relative difference $\Delta \mathrm{AIC}$ with respect to the best-fit model (lowest AIC). Rather than selecting a single best-fit solution, we considered all models with $\Delta \mathrm{AIC} < 4$ as statistically indistinguishable. Those with $\Delta\mathrm{AIC} > 10$ are strongly disfavored and can be excluded \citep{Burnham}. In Figure \ref{fig:barplot} we show the result of the relative fraction of column density for each compound. The fractions were calculated using models with $\Delta \mathrm{AIC} < 4$. The symbol $a$ represents amorphous compounds, while $c$ denotes crystalline compounds. The error bars indicate the standard deviation of each fraction within the $\Delta \mathrm{AIC} < 4$ selected models. Figure \ref{fig:FeK} presents a zoom-in on the Fe K-edge along with one selected best-fit model, and the fit residuals. The bottom panel of this plot shows the transmission of the gas and dust components.

%%%%%% Dust samples Table %%%%%%%%%%
\begin{table}
\caption{List of dust extinction models used for the Fe K-edge modeling. A detailed description of the models is provided in \cite{Rogantini2018}.}
\centering
\begin{tabular}{cccc}
  \hline
  \hline
 \# & Compound & Chemical Formula & Form \\
  \hline
  1 &  Olivine & $ \rm Mg_{1.56}Fe_{0.4}Si_{0.91}O_{4}$ &  crystalline \\
 2 &  Pyroxene & $ \rm Mg_{0.6}Fe_{0.4}SiO_{3}$ &  amorphous \\
 3 &  Pyroxene &  $ \rm Mg_{0.6}Fe_{0.4}SiO_{3}$ &  crystalline \\
 4 &  Pyroxene & $ \rm Mg_{0.9}Fe_{0.1}SiO_{3}$ &  amorphous \\
 5 &  Pyroxene &  $ \rm Mg_{0.9}Fe_{0.1}SiO_{3}$ &  crystalline \\
 6 & Pyroxene & $ \rm Mg_{0.75}Fe_{0.25}SiO_{3}$ &  crystalline \\
 7 &  Troilite & $ \rm FeS$ &  crystalline\\
 8 &  Pyrrhotite & $ \rm Fe_{7}S_{8}$ &  crystalline\\
 9 &  Metallic iron\footnote{The metallic iron model is based on the laboratory measurements of \citet{Lee2005}.
} & $ \rm Fe$ &  -         \\
\hline
\hline
\end{tabular}

\label{tab:samples}
\end{table}

\section{Discussion} \label{Discussion}

\subsection{The best fit dust compound}

We find that the total dust column density from our fit is distributed among three main components: $\sim 20\%$ in metallic iron, $\sim 27\%$ in iron sulfides, and the rest in Mg-rich silicates. These values represent the fractions of the total dust column density contributed by each class of compounds included in the fit.
These results should, however, be interpreted with some caution. According to the Akaike Information Criterion, approximately $25\%$ of the tested models reproduced the Fe K-edge equally well, which means that it is challenging to distinguish one best fit dust combination. 
Nevertheless, valuable knowledge can still be gained from the overall column density fractions, as certain compounds consistently appear as major contributors across the best-fit models, revealing a trend.

This result is broadly consistent with our previous X-ray studies of the ISM. For instance, studies of the Fe L-edges show that while silicates dominate the dust grain chemistry result, a significant fraction of metallic iron is also present, contributing up to 40\% to the total dust mass \citep{Psaradaki2022, Psaradaki2026}. These results could further support the view that interstellar silicates are predominantly magnesium-rich, and that the silicate mixture dust composition lies between pyroxene and olivine \citep{Min2007, Molster2002, Chiar2006}.

\citet{Rogantini2026} studied the Chandra High Energy Transmission Grating (HETGS) spectrum of GX~340+0, focusing on the dust absorption through a simultaneous fit of the Si K, Fe K, and S K edges. 
They find that amorphous olivine ($\mathrm{MgFeSiO_4}$) provides the dominant contribution to the dust column density ($\sim 65\%$), followed by metallic iron ($\sim$19\%) and iron sulfides (FeS and Fe$_8$S$_7$; $\sim$10\%). Our fraction of metallic iron is consistent with their result; however, we find a larger contribution from iron sulfides in the XRISM data. Furthermore, while \citet{Rogantini2026} identify amorphous olivine as the dominant silicate component, we do not include it in our Fe K-edge analysis because our database currently provides an equivalent model for amorphous olivine only at the Si K- and Mg K-edges, but not at the Fe K-edge. Instead, we find a significant contribution to the dust mass from silicates with a lower Fe content and a higher Mg fraction compared to amorphous olivine.

These differences are likely related to the different datasets and modeling approaches adopted in the two studies. In this work, we focus exclusively on the Fe K-edge using XRISM data, which provides improved spectral resolution in this energy range compared to Chandra, but does not include information from the Si K-edge. As discussed by \citet{Rogantini2026}, the preference between different dust models is strongly driven by the Si K-edge, which has the highest signal-to-noise ratio and exhibits the strongest X-ray absorption fine structure (XAFS) signatures. A simultaneous analysis of the Fe K and Si K edges using the XRISM and Chandra datasets could therefore provide stronger constraints on the dust composition and help resolve the remaining uncertainties. We leave this approach for future investigation.

\subsection{The reservoirs of iron}

Identifying the reservoirs of solid-phase iron remains an active area of study. Here we quantify how the iron atoms are distributed among the different dust species. These fractions differ from the dust composition fractions shown in Figure~\ref{fig:barplot}, which give the share of the total dust mass in each compound. Compounds that dominate the dust mass but are relatively poor in iron (e.g. silicates) can therefore contribute much less to the iron budget than to the dust mass. To obtain the iron fractions, we consider the iron mass contained in each compound, using all the selected models with $\Delta{\rm AIC}<4$. We find that approximately 38\% of the iron is in metallic form, 47\% in iron sulfides, and the remaining in silicates.

\citet{Draine_2021} suggested a relatively small metallic iron fraction ($\lesssim 10\%$), based on constraints from the thermal magnetic-dipole emission expected from ferromagnetic grains \citep{Draine_2013}. However, \citet{Hensley_2023} suggested that these emission estimates may be overpredicted, leaving open the possibility that a substantially larger reservoir of metallic iron could be present in interstellar dust. Consequently, the role of metallic iron as a major iron-bearing dust component is an active area of study. 

\citet{Dwek1998} discuss that Type Ia supernovae, the primary sources of iron in the ISM, produce more iron than can be incorporated into oxides and sulfides only. As a result, a significant fraction of the iron may condense in the form of metallic iron grains. Moreover, \citet{Poteet} analyzed interstellar grains toward $\zeta$ Ophiuchi, an interstellar environment near the diffuse-dense cloud transition, and found that nearly all Mg and Si atoms along the line of sight reside in amorphous silicate grains, while a substantial fraction of elemental Fe is found in compounds other than silicates. Similarly, \citet{Westphal} suggested that X-ray observations along the Cygnus X-1 sightline are consistent with a significant amount of Fe being in metallic form. \citet{Zhukovska2018}, concluded that the majority ($\sim$70\%) of interstellar iron resides in metallic nanoparticles, and \cite{Kohler2014} showed that metallic Fe inclusions embedded in amorphous silicate grains can reproduce the observed dust properties. Together with our results, these suggest that metallic iron may constitute a significant reservoir of Fe in the ISM.

% sulfites
The fraction of sulfur locked into dust grains, as well as the amount of Fe included into sulfides, remain uncertain. The results suggest that a significant fraction of iron residing in sulfite compounds, in the form of troilite (FeS) and pyrrhotite (Fe$_7$S$_8$).
\citet{Westphal} suggested that Fe is approximately equally partitioned between metallic iron and sulfides, consistent with a high-temperature condensation sequence in which both Fe and S are predominantly incorporated into these phases. They also report that less than 38.5\% ($2\sigma$) of Fe resides in amorphous silicates, which is consistent with our study.

\citet{Corrales2025}, showed that Fe sulfides provide a good description of the S K-edge using the same XRISM observations of GX~340+0 analyzed in this work. From their spectral fit, they derived an upper limit of 25\% on the fraction of interstellar Fe bound in Fe--S compounds. Our Fe K-edge fit suggests a somewhat larger contribution from these compounds. Overall, these results support the presence of Fe sulfides in the ISM, although their precise contribution to the total iron budget needs to be confirmed. A joint analysis of both absorption edges could, possibly, provide stronger constraints on the abundance of Fe sulfides in the ISM.

% X-rays
Our results can also be placed in the context of previous X-ray studies of interstellar iron. \cite{Rogantini2020}, by analyzing multiple X-ray edges of several bright binaries using Chandra/HETGS data, found that the major iron reservoir is in silicates using the Si K and Mg K edges. In \citet{Psaradaki2022}, we analyzed the Fe L-edges of five diffuse sightlines in the Galaxy and found that, on average, about 40\% of dust-phase iron is metallic, with the remaining 60\% incorporated into silicates. In a more recent study in \cite{Psaradaki2026}, we examined the line of sight to GX 339–4, which has a moderate Galactic column density, placing it between dense and diffuse sightlines. There, roughly 65\% of iron is metallic, 32\% is in silicates, and 3\% is in other compounds such as FeS.

Comparing our results with previous X-ray studies suggests that the fraction of iron residing in metallic iron and sulfides may vary between different Galactic sightlines. With the Fe K-edge we probe a higher column density interstellar medium, compared to other studies, for example with the Fe L-edges which probes diffuse ISM. The higher fraction of Fe sulfides inferred here could reflect differences in the ISM environments probed by these observations. In fact, sulfur is known to become increasingly depleted from the gas phase in high-column density regions, suggesting that sulfur-bearing dust species may represent a more significant reservoir in denser environments \cite[e.g.][]{Sofia1994, Savage1996, Howk2006, Jenkins2009, Woods2015}. 

A possible limitation of this study comes from the different photoabsorption edges used in the literature and from the dust models available for each photoabsorption edge. Different edges may have different sensitivities to specific iron-bearing compounds, which can affect the derived iron fractions. For example, \citet{Rogantini2026}, who studied the same source using Chandra data, discuss that the Si K-edge could be mainly responsible for constraining the dust composition, with amorphous olivine ($\rm MgFeSiO_{4}$) providing the best fit. However, this compound is not included in the Fe K-edge models used in this work, which may introduce additional uncertainty in our results. The Fe K-edge spectral shape (at least for GX 340+0) may also be more sensitive to certain iron-bearing species than the Fe L-edge, potentially leading to differences in the inferred contribution of metallic iron and iron sulfides. Future studies combining multiple absorption edges, together with improved laboratory measurements and deeper XRISM observations, will be needed to constrain more accurately the distribution of iron among its different forms in the interstellar medium.

Up-to-date laboratory measurements of neutral iron species, particularly solid metallic iron, atomic Fe, and iron sulfides are essential to better constrain the fraction of Fe present in these minerals. Metallic iron, has not been recently studied in the laboratory due to challenges with oxidation. This highlights the need for new laboratory experiments. This work is ongoing, and we anticipate that improved laboratory data and dust extinction models for metallic iron will become available in the coming years, enabling more robust comparisons between different observations.
Laboratory data for other Fe K-edge compounds, including amorphous olivine, as well as S K-edge compounds, have already been acquired and will be incorporated in a future study (S. Zeegers, private communication).
An additional source of uncertainty concerns the ionization state of the gas-phase iron along the line of sight. UV and optical observations indicate that in neutral regions of the ISM, most gas-phase iron exists as \feii, because \fei is easily ionized by photons from the interstellar radiation field, while \feii is much less susceptible due to its higher ionization potential (\citealt{Snow2002}; \citealt{Miller2007}; \citealt{Jensen2007}). In H II regions, gas-phase iron is expected to be a mixture of \feii, \feiii, and \feiv. Any remaining \fei is likely locked in dust grains, given depletion rates of up to 99\% in cold neutral regions \citep[e.g.,][]{Savage1996}. In \citet{Psaradaki2024}, we analyzed combined X-ray and UV data for a Galactic X-ray binary probing the diffuse ISM and found that the majority of gas-phase iron is in the form of \feii. 
% !!! %{\color{red} say why you concluded this -- For the present study, however, the gas-phase iron appears to be predominantly neutral \fei.} 
Since no clear \feii feature is detected in the X-ray data, future work could include UV spectroscopy along the line of sight to GX 340+0 to assess whether in this denser environment \feii can be reliably constrained.

\subsection{The abundance and depletion of Fe}

\begin{table}
\caption{Iron column densities and abundances.}
\label{tab:depletions}      % is used to refer this table in the text
\centering
\renewcommand{\arraystretch}{1.1}%                     % used for centering table
\begin{tabular}{cc}        % centered columns (4 columns)
\hline           % inserts double horizontal lines
\hline
$N_{\rm gas}$   & $(1.49\pm0.85)\times10^{17}\ {\rm cm^{-2}}$ \\
$N_{\rm dust}$  & $(2.17\pm0.31)\times10^{18}\ {\rm cm^{-2}}$ \\
$A_{\rm gas}$   & $(2.16\pm1.80)\times10^{-6}$ \\
$A_{\rm dust}$  & $(3.15\pm1.97)\times10^{-5}$ \\
$A/A_\odot$     & $1.03\pm0.60$ \\
$\delta_{X}$   & $0.94^{+0.06}_{-0.58}$ \\
\hline 
\hline
\end{tabular}
\begin{flushleft}
\footnotesize{$N_{\rm gas}$ and $N_{\rm dust}$ indicate the total column density of gas and dust, in $\rm cm^{-2}$, respectively. $ A_{\rm gas}$ and $A_{\rm dust}$ are the abundances of iron in gas and dust respectively (with respect to hydrogen). $ A/A_\odot$ is the total abundance (gas+dust) ratio in proto-Solar abundance units, and $\delta_{X}$ is the depletion of this element into solids.}
\end{flushleft}
\end{table}

Table \ref{tab:depletions} summarizes the calculated elemental abundances and depletions for Fe. $N_{\rm gas}$ and $N_{\rm dust}$ denote the total column densities of gas and dust, respectively. $A_{\rm gas}$ and $A_{\rm dust}$ represent the abundances of iron in the gas and dust phases, relative to hydrogen. The quantity $A/A_\odot$ corresponds to the total (gas + dust) abundance, while $\delta_{X}$ is the depletion of the element into solid form. We find that Fe is predominantly depleted into dust, as it has been found from the Fe L-edges studies of the diffuse ISM studies \cite[e.g.][] {Costantini2012,Pinto2013, Psaradaki2022, Corrales2024}. This degree of depletion agrees with \citet{Jenkins2009}, who examined more than 200 sightlines within 3 kpc of the Milky Way and found that no more than 10\% of iron remains in the gas phase.

In this work, we find the iron abundance to be consistent with the Solar value reported by \citet{Lodders2009}. However, given the relatively large uncertainties associated with this measurement, this result should be interpreted with caution. A potential slight super-Solar iron abundance toward GX 340+0 is expected, as this sightline probes the inner Galaxy, where higher metallicities are presumed due to the Galactic abundance gradient \citep{Dwek1998}. 

A slight increased Fe abundance compared to Solar would be consistent with the results of \citet{Corrales2024}, who studied GX 339–4, as well as with our recent analysis of the same source using Chandra observations \citep{Psaradaki2026}. Moreover, \citet{Gatuzz2016} studied a large sample of sources along the Galactic plane and found that the iron abundance varies between about 0.1 and 2 times the Solar value within the errors (adopting the abundance scale of \citealt{Grevesse1998}), depending on the line of sight. This range is fully consistent with our measurement uncertainties, even when accounting for the differences between abundance tables (see \citealt{Psaradaki2024} for a detailed comparison of commonly used abundance tables).

Deeper XRISM observations of the Fe K-edge in multiple Galactic sightlines, together with improved laboratory measurements, will be essential for confirming these results and improve our understanding of the interstellar iron budget.

\section{Conclusions}

In this study, we present a pilot high-resolution X-ray investigation of the Fe K-edge and Fe-bearing dust grains in a high column density Galactic sightline using XRISM/Resolve observations of GX~340+0. Our main conclusions, for this sightline, are the following:

\begin{itemize}

\item The dust column toward GX~340+0 is best described by a mixture of silicates ($\sim$53\%), metallic iron ($\sim$20\%), and iron sulfides ($\sim$27\%). Although the Fe K-edge does not provide a unique dust composition, these components consistently appear among the statistically acceptable models, indicating that they could be significant contributors to the dust mass towards GX 340+0.

\item The iron budget derived from the dust models indicates that $\sim$38\% of solid-phase Fe is contained in metallic iron, $\sim 47\%$ in iron sulfides, and the remainder in silicates. The large contribution from metallic iron and iron sulfides differs from what is typically inferred from X-ray studies of diffuse ISM sightlines and may support the evidence that that these compounds become more significant reservoirs of Fe in higher column density environments.

\item We find that iron is strongly depleted from the gas phase, with approximately $95\%$ of Fe residing in dust. This result could support the possibility that metallic iron and iron sulfides represent important reservoirs of interstellar solid-phase Fe. 

\item Updated laboratory measurements of neutral Fe-bearing compounds at the Fe K-edge, particularly solid metallic iron, atomic \fei, and iron sulfides, are required to better constrain the fraction of Fe residing in these phases.

\item Given that the gate valve is closed, deeper XRISM/Resolve observations will be important to place stronger constraints on the iron abundance and depletion pattern, and to more robustly determine the preferred dust mineralogy through the Fe K-edge.

\end{itemize}

We thank the referee for the helpful comments. IP would like to thank Renee Ludlam for useful discussions on XRISM data reduction. 
IP acknowledges support through the European Space Agency (ESA) Research Fellowship in Space Science, and thanks SRON (Space Research Organisation Netherlands) for their hospitality.
This work was also supported by the grant 80NSSC24K0230 for the XRISM PV observations as part of NASA's Guest Scientist program.
 Additional support for this work was provided by NASA through the Smithsonian Astrophysical Observatory (SAO) contract SV3-73016 to MIT for Support of the Chandra X-Ray Center (CXC) and Science Instruments. CXC is operated by SAO for and on behalf of NASA under contract NAS8-03060. PC acknowledges support from NASA XRISM grant 80NSSC23K0637 and NASA Chandra grants GO3-24024X 16619349, and GO4-25094X 16619354.

\bibliography{reference}{}
\bibliographystyle{aasjournal}

\end{document}